\documentclass[12pt]{article}
\usepackage[english]{babel}
\usepackage[dvips]{graphicx}

\title{Extra quark-lepton generations and precision measurements}
\author{M.Maltoni \\
Dipartimento di Fisica, Universit\'{a} di Ferrara, I-44100, Ferrara, Italy
\\
Istituto TeSRE/CNR, Via Gobetti 101, I-40129 Bologna, Italy \\
INFN, Sezione di Ferrara, I-44100, Ferrara, Italy \\
mmaltoni@fe.infn.it \\
~~ \\
V.A.Novikov, L.B.Okun \\
ITEP, 117218, Moscow, Russia \\
novikov@heron.itep.ru \,  okun@heron.itep.ru \\
  \\
A.N.Rozanov \\
ITEP and CPPM, IN2P3-CNRS, Univ. M\'editerran\'ee, F-13288,  \\
Marseilles, France \\
rozanov@cppm.in2p3.fr  \\
~~ \\
M.I.Vysotsky \\
ITEP and INFN, Sezione di Ferrara, Italy  \\
vysotsky@heron.itep.ru}
\date{29.11.99}

\def\ga{\mathrel{\mathpalette\fun >}}
\def\fun#1#2{\lower3.6pt\vbox{\baselineskip0pt\lineskip.9pt
\ialign{$\mathsurround=0pt#1\hfil##\hfil$\crcr#2\crcr\sim\crcr}}}

\begin{document}
\maketitle

\begin{abstract}
The existence of extra chiral generations with all 
fermions heavier than $M_Z$
is strongly disfavoured by the precision
electroweak data. However the data are fitted nicely even by a few extra
generations, if
one allows neutral leptons to have masses close
to $50$ GeV.
The  data allow inclusion of
one additional generation of heavy fermions in SUSY extension of Standard
Model 
if chargino and
neutralino have masses close to $60$ GeV with  $\Delta m \simeq 1$ GeV.
\end{abstract}

\newpage

\section{Introduction}

The aim of this article is to analyze to what extent 
the existing data on the $Z$-boson
parameters, the $W$-boson and the top quark masses
allow to bound effectively New Physics at high  energies
which does not decouple at ``low"
($\sim m_{Z}$) energies. The most straightforward generalization of the
Standard Model (SM) through inclusion of extra chiral generation(s) of
heavy fermions, quarks ($q=U,D$)  and leptons ($l=N,E$), is an example of such
non decoupled New Physics. We show that it is excluded by 
the electroweak precision data, if all
extra fermions are heavy: $m \ga m_Z$. 
We find however that if  masses of new neutrinos are close to $50$ GeV,
additional generations become allowed: even three extra generations with
$\approx 50$ GeV neutrinos can exist.

Finally inclusion of new generations in SUSY extension of Standard Model is
discussed.

When speaking on extra generations the first thing to bother about is the
extra neutrinos, $N$.
Being coupled to $Z$-boson they would increase the invisible $Z$-width.
To avoid
contradiction with experimental data their masses should be larger than
$45$ GeV \cite{1X}. In order to meet this condition one has to introduce
not only  left-handed neutrinos $N_L$, but also right-handed neutrinos
$N_R$ and supply new ``neutrinos" with Dirac masses
analogously to the case of charged leptons and quarks. Unfortunately gauge
symmetries do not forbid Majorana mass for $N_R$, and if it is large one
would get light $N_L$ through the see-saw mechanism.  To avoid such a failure
we will suppose that the Majorana mass of $N_R$ is negligible, closing eyes
on the emerging (un)naturalness problem. Thus, our neutral lepton $N$
is a heavy Dirac particle.

Contribution of new generations to electroweak radiative corrections was
considered in  papers \cite{2X} -  \cite{33}.
In what follows we will assume that the mixing among new generations and
the three existing ones is small, hence new fermions contribute only to
oblique corrections (vector boson self energies). This case is
discussed  in \cite{33} and the authors come to the conclusion that one extra
generation of heavy fermions is excluded at 99.2\% CL.
The authors of \cite{33} follow a two step
procedure: first they find that experimental bounds on parameter $\rho$
exclude existence of nondegenerate extra generation
(degenerate generation is decoupled from $\rho$). Then they consider
 parameter
$S$  \cite{111} from which degenerate generation is not decoupled
and find that
experimental value of $S$ excludes the existence of extra degenerate
generation as well. Such procedure is not general enough: it does not use
all precision data.
In the present paper we perform global fit of all precision data.

 We study both degenerate and
nondegenerate extra generations on
the equal footing. By considering the contributions of new generations into
all precision electroweak
observables simultaneously we see that the fit of the data is worsened
by them if all new particles are heavy. Taking the number
of new generations $N_g$ as a continuous parameter
(just as it was done with the
determination of the number of neutrinos from invisible $Z$ width) we
get a bound on it. The $\chi^2$ minimum corresponds to $N_g \simeq
-0.5$, while $N_g =1$ is excluded by more than 2 standard deviations.

Section 2 contains general formulas for oblique radiative corrections 
caused  by an extra doublet of  quarks or leptons. In section 3 we consider
the  case when all extra fermions are heavy ($m \ga m_Z$). In section 4 we 
consider the case when $U,D,E$ are heavy, while $N$ is ``light'',
$m_N \simeq 50 GeV$. In this case the  contribution of $N$ compensates 
that of $U,D,E$. Section 5 analyzes the SUSY version of four  generations.

\section{Formulas}

New particles contribute to physical observables through
self-energies
 of vector and axial currents. This gives
corrections $\delta V_i$
to the functions $V_i (i=A,R,m)$ which determine the values of
physical
observables (axial coupling $g_A$, the ratio $R=g_V/g_A$, and the ratio
$m_W/m_Z$) \cite{2}:

\begin{equation}
\frac{3\bar{\alpha}}{16\pi s^2 c^2}\delta V_A =
\Pi_Z(m_Z^2)-\Pi_W(0)-\Sigma^{\prime}_Z(m_Z^2) \;\; ,
\label{100}
\end{equation}

\begin{equation}
\frac{3\bar{\alpha}}{4\pi}\delta V_R =
4c^2 s^2[\Pi_Z(m_Z^2)-\Pi_W(0)-\Sigma^{\prime}_{\gamma}(0)]-
4cs(c^2 -s^2)\Pi_{\gamma Z}(m_Z^2) \;\; ,
\label{101}
\end{equation}

\begin{equation}
\frac{3\bar{\alpha}}{16\pi s^4}\delta V_m =
\frac{c^2}{s^2}\Pi_Z(m_Z^2)+\frac{s^2
-c^2}{s^2}\Pi_W(m_W^2)-\Pi_W(0)-\Sigma_{\gamma}^{\prime}(0) \;\; ,
\label{102}
\end{equation}
where $\bar{\alpha}\equiv \alpha(m_Z^2)$ -- the value of electromagnetic
coupling at $q^2 = m_Z^2$; $s\equiv\sin\theta$,
$c\equiv \cos\theta$, $\theta$ being the electroweak mixing angle,
$s^2=0.23116(23)$; $\Pi_i (q^2)=\Sigma _i (q^2)/m_i^2, i=W,Z$;
$\Pi _{\gamma Z}=\Sigma _{\gamma Z}/m_Z^2$; $\Sigma^{\prime}(0)=
 \lim_{q^2 \rightarrow 0}\Sigma(q^2) /q^2$, where $\Sigma$ is a vector boson
self-energy (see \cite{2}).

In this section we present expressions for these
corrections. We start with the case of one $SU(2)_L$ doublet and its
two right-handed singlet
companions: $(UD)_L$, $U_R$, $D_R$. Let $N_c^{q,l}$ be the number of
colors ($N_c^{q}=3, N_c^{l}=1$) and $u\equiv m_U^2/m_Z^2$,
$d\equiv m_D^2/m_Z^2$.
$Y^{q,l}$ is the doublet hypercharge, $Y^{q,l}=Q_U +Q_D$,
where $Q_U$ and $Q_D$ are
$U$ and
$D$ electric charges (the hypercharge of isosinglets being equal to its
doubled electric charge).
Corrections produced by the one generation fermions are equal to the sum of
lepton and
quark contributions:

\begin{equation}
\Delta V_i=\delta V^q_i +\delta V^l_i
\end{equation}

Contributions of quark or lepton doublets are given by:

\begin{eqnarray}
\delta V_A^{q,l} &=& \frac{N_c^{q,l}}{3}\left[ u+d +
\frac{2ud}{d-u} \ln(\frac{u}{d})
-F^{\prime}(u) -F^{\prime}(d)\right] + \nonumber \\
&+&N_c^{q,l}(-\frac{4}{9}s^2 -\frac{1}{9})\left[2uF(u) -(1+2u) F^{\prime}(u) +
2dF(d)
-(1+2d)F^{\prime}(d) \right] + \nonumber \\
&+&\frac{16}{9}N_c^{q,l} s^4\left\{Q_U^2\left[2uF(u) -(1+2u)
F^{\prime}(u)]+Q_D^2[2dF(d) -(1+2d)F^{\prime}(d) \right]\right\} +
\nonumber \\
&+&\frac{4}{9}N_c^{q,l} s^2 Y^{q,l}\left[2dF(d) -(1+2d)
F^{\prime}(d)-2uF(u) +(1+2u)F^{\prime}(u) \right] \;\; ,
\label{1}
\end{eqnarray}

\begin{eqnarray}
\delta V_R^{q,l} &=& -\frac{2N_c^{q,l}}{3}\left\{ uF(u)+dF(d) +\frac{ud}{u-d}
\ln(\frac{u}{d}) -\frac{1}{2}(u+d)\right\} + \nonumber \\
&+&\frac{16}{9}N_c^{q,l} s^2 c^2\left\{Q_U^2[(1+2u)
F(u)-\frac{1}{3}]+Q_D^2[(1+2d)F(d)-\frac{1}{3}]\right\} + \nonumber \\
&+&\frac{2N_c^{q,l} Y^{q,l}}{9}\left\{(1+2d) F(d)-(1+2u)F(u)+\ln(\frac{u}{d})
 \right\} \;\; ,
\label{2}
\end{eqnarray}

\begin{eqnarray}
\delta V_m^{q,l} &=& \frac{2}{9}N_c^{q,l}(1-\frac{s^2}{c^2})\left\{-F
(m_W^2, m_U^2,
m_D^2)[2c^2 -u-d-\frac{(u-d)^2}{c^2}]+u+d-\frac{4}{3}c^2\right\} - \nonumber
\\
&-&\frac{4s^2}{9}N_c^{q,l} Y^{q,l}\left[(1+2u) F(u)
-(1+2d)F(d)-\ln(\frac{u}{d}) \right] +  \nonumber \\
&+&\frac{16}{9}N_c^{q,l} s^4\left[Q_U^2\left((1+2u)
F(u)-\frac{1}{3}\right)+Q_D^2\left((1+2d)F(d)-\frac{1}{3}\right)\right]
+\frac{s^2}{3c^2}N_c^{q,l}(u+d)+ \nonumber \\
&+&\frac{2}{9}N_c^{q,l} \left[(1-u) F(u)+(1-d)F(d)-\frac{2}{3}\right]-
\nonumber \\
&-&\frac{4}{9}N_c^{q,l} s^2\left[(1+2u)F(u)+(1+2d)F(d)-\frac{2}{3}\right]
+ \nonumber
\\
&+&\frac{2}{9}N_c^{q,l} \ln(\frac{u}{d})\left[(1+\frac{1}{c^2})
\frac{ud}{d-u}+(s^2
-c^2)\frac{d+u}{d-u} \right] \;\; ,
\label{3}
\end{eqnarray}
where
\begin{eqnarray}
F(m_W^2, m_U^2, m_D^2)&=&-1+\frac{m_U^2 +m_D^2}{m_U^2 -m_D^2}
\ln(\frac{m_U}{m_D}) - \nonumber \\
&-&\int\limits^1_0 dx \ln\frac{x^2 m_W^2 -x(m_W^2 +m_U^2 -m_D^2) +m_U^2}{m_U
m_D} \;\; ,
\label{4}
\end{eqnarray}

\begin{eqnarray}
F(u)=F(m_Z^2, m_U^2, m_U^2) = \left\{
\begin{array}{ll}
2\left[1-\sqrt{4u-1}\arcsin(\frac{1}{\sqrt{4u}})\right] \; , & u >
\frac{1}{4} \\
2\left[1-\sqrt{1-4u}\ln\left(\frac{1+\sqrt{1-4u}}{\sqrt{4u}}\right)\right] \;
, & u < \frac{1}{4}
\end{array}
\right.
\label{5}
\end{eqnarray}

\begin{equation}
F^{\prime}(u) \equiv -u\frac{d}{du}F(u) =\frac{1-2u F(u)}{4u -1} \;\; .
\label{6}
\end{equation}

The following relation is useful in deriving eqs (5) - (7):
\begin{equation}
\int\limits^1_0 dx(x^2 -x)\ln(x^2 -x+u) =\frac{1+2u}{6}F(u) -\frac{1}{18}
-\frac{1}{6}\ln u \;\; .
\label{112}
\end{equation}

In the asymptotic limit (denoted by prime) where the fourth generation
particles are much heavier than
$Z$-boson ($u, d\gg 1$) neglecting power suppressed terms we obtain
from eqs (5) - (7):
\begin{equation}
\delta^{\prime}V_A^{q,l} = \frac{N_c^{q,l}}{3} [u+d+\frac{2ud}{d-u}
\ln(\frac{u}{d})] \;\;
,
\label{7}
\end{equation}
\begin{equation}
\delta^{\prime}V_R^{q,l} =\delta^{\prime}V_A^{q,l}
-\frac{2}{9}N_c^{q,l} +\frac{2}{9}N_c^{q,l} Y^{q,l}\ln(\frac{u}{d}) \;\; ,
\label{8}
\end{equation}

\begin{eqnarray}
\delta^{\prime}V_m^{q,l} &=& \delta^{\prime}V_A^{q,l}
-\frac{2}{9}N_c^{q,l} +\frac{4}{9}N_c^{q,l}s^2 Y^{q,l} \ln(\frac{u}{d}) +
\frac{4}{9}N_c^{q,l} (s^2-c^2) \times    \nonumber \\
&\times&\left[\frac{1}{3}-\frac{2}{(\frac{u}{d}-1)^2}
\frac{u}{d} +\frac{3}{(\frac{u}{d}-1)^3}(\frac{u}{d}-\frac{1}{3}
-\frac{1}{6}(\frac{u}{d}-1)^3) \ln(\frac{u}{d})\right] \;\; .
\label{9}
\end{eqnarray}

In order to obtain the contribution of a generation
 one should sum those of
quarks and leptons:

\begin{equation}
\Delta^{\prime} V_i=\delta^{\prime} V^q_i +\delta^{\prime} V^l_i
\end{equation}

In  the case of fully degenerate doublets ($m_U=m_D=m_N=m_E$) one obtains from
(\ref{7}) - (\ref{9}):
\begin{equation}
\Delta^{\prime}V_A = 0 \; , \;\; \Delta^{\prime}V_R
=-\frac{8}{9} \; , \;\; \Delta^{\prime}V_m =-\frac{16}{9}s^2 \;\; .
\label{10}
\end{equation}

In the opposite case when $SU(2)_V$ is strongly violated  ($m_U \gg m_D$
or $m_U \ll m_D$)     all
corrections $\Delta^{\prime}V_i$ become equal:
\begin{equation}
\Delta^{\prime}V_i =\frac{|m_U^2 -m_D^2|}{m_Z^2} + \frac{1}{3}\frac{|m_N^2
-m_E^2|}{m_Z^2} \;\; .
\label{11}
\end{equation}

 In the present paper we consider the general case described not by
asymptotic formulas (\ref{7}) - (\ref{9}) but by general formulas
(\ref{1}) - (\ref{3}) which will allow us to study the case of "light" new
neutrinos ($m_N \approx m_Z/2$).  Using formulas (\ref{7}) - (\ref{9}) we
already analyzed the effect of new generations with all fermions heavy,
$m \ga m_Z$, in  paper \cite{3} (see also
\cite{34},\cite{35}).  However with new experimental data
we get
much more restrictive bounds than those obtained in \cite{3} (see the next
section).

Let us present at the end of this section formulas for the ``horizontally
degenerate" case ($m_{U} = m_{N}$, $m_{D} = m_{E}$):
\begin{eqnarray}
\Delta V_A&=&
\frac{4}{9}\left\{(\frac{16}{3}s^4 -4s^2 -1)[2u F(u)-(1+2u)F'(u)+2dF(d)
-(1+2d)F'(d)] + \right.
\nonumber \\
&+& \left. 3[u+d -\frac{2ud}{u-d}\log\frac{u}{d} -F'(u) -F'(d)]\right\} \;\; ,
\label{12}
\end{eqnarray}

\begin{equation}
\Delta V_R =
-\frac{8}{3}[u F(u)+dF(d)+\frac{ud}{u-d}\log\frac{u}{d} -\frac{u+d}{2}]
 +\frac{64}{27}s^2 c^2 [(1+2u)F(u)+(1+2d)F(d)
-\frac{2}{3}] \;\; ,
\label{13}
\end{equation}

\begin{eqnarray}
\Delta V_m &=&
(\frac{64}{27}s^4 -\frac{16}{9}s^2)[(1+2u)F(u)+(1+2d)F(d)-\frac{2}{3}] +
\nonumber \\
&+& \frac{8}{9}[(1-u)F(u)+(1-d)F(d)-\frac{2}{3}]+\frac{4}{3}\frac{s^2}{c^2}
[u+d-\frac{2ud}{u-d}\log\frac{u}{d}] +
\nonumber \\
&+& \frac{8}{9}(1-\frac{s^2}{c^2})[\frac{u-d}{2}\log\frac{u}{d}+(u+d)+(c^2
-\frac{u+d}{2})\frac{u+d}{u-d}\log\frac{u}{d}-\frac{4}{3}c^2 -
\nonumber \\
&-&(2c^2 -u-d-\frac{(u-d)^2}{c^2})F(m_W^2, m_U^2, m_D^2)] \;\; .
\label{14}
\end{eqnarray}

 In Fig. 1 the $u$ dependence
of the functions
$\Delta V_i$ and $\Delta^{\prime} V_i$  for $d=1$ is
shown. It is
clear that accuracy of equations (\ref{7}) - (\ref{9}) is very good as soon
as new fermions are heavier than $Z$-boson.

\section{Comparison with experimental data: heavy fermions}

We compare theoretical predictions for the case of
the presence of extra generations with
experimental data with the help of the code LEPTOP \cite{9XX}
(see also \cite{2}). We take
$m_D = 130$ GeV --
the lowest value allowed for the new quark mass from Tevatron search
\cite{13}  and take
$m_U \ga m_D$.
As for the leptons from the extra generations, their masses are independent
parameters. To simplify the analyses we start with
 $m_N = m_U$, $m_E = m_D$.
Any value of higgs mass above $90$ GeV is allowed in our fits, however
$\chi^2$ appears to be minimal for $m_H=90$ GeV.  In Figure 2 the excluded
domains in coordinates ($N_g$, $\Delta m$) are shown 
(here $\Delta m = (m_U^2 - m_D^2)^{1/2}$).  Minimum
  of $\chi^2$ corresponds to $N_g =-0.5$ and the case $N_g=0$ is in the one
standard deviation domain ($1\sigma$).  We see that one extra generation
corresponds to $2.5\sigma$ approximately. The behaviour of division lines in
Fig. 2 can be understood qualitatively.  For degenerate extra generations the
corrections $\Delta V_i$ are negative.  They become positive and large when
$\Delta m$ increases. That is why at large $\Delta m$ division lines approach
$N_g=0$ value. In the intermediate region ($\Delta m \approx 125$ GeV)
$\Delta V_i$ cross zero and this explains the turn to the right of the
division lines. However, for different $i$ zero is reached for different
$\Delta m$ values, that is why extra generations are excluded even for
$\Delta m \approx 125$ GeV (see \cite{3}).

We checked that similar bounds are valid for the general choice of heavy
masses of leptons and quarks. In particular we found that for 
 $m_N = m_D = 130$ GeV and $m_E = m_U$ one extra generation is excluded
at 2 $\sigma$ level, while for  $m_E = m_U = 130$ GeV and $m_N = m_D$
the limits are even stronger than in Fig.2.

\section{Comparison with experimental data: $m_N < m_Z$}

According to \cite{13} lower bound on $m_E$ from LEP II is approximately $80$
GeV.  However, quasi-stable neutral lepton $N$ can be considerably lighter.
From LEP II searches of the decays $N\to lW^{\ast}$, where $W^{\ast}$ is
virtual while $l$ is $e$, $\mu$ or $\tau$, it follows that $m_N > 70-80$ GeV
for the mixing angle with three known generations larger than $10^{-6}$
\cite{6X}.  Thus let us take in this section this mixing to be less than
$10^{-6}$. In this case only DELPHI bound from the measurement of the
$Z$-boson width is applicable, $m_N > 45$ GeV \cite{1X}. If $m_N$ is larger
than $m_Z/2$ searches at LEP II of the reaction $e^+ e^- \to N \bar{N} \gamma$
should
bound $m_N$. The observation of a ``lonely photon''
 was suggested long time ago as a method to study
cross section of the $e^+e^-$ annihilation into neutrinos \cite{11X}. DELPHI
collaboration performed such a search at $E\leq 183$ GeV and found that total
number of neutrinos $N_{\nu} = 2.92 \pm 0.25 {(\rm stat)} \; \pm 0.14 {\rm
(syst)}$ \cite{10X}.  However most of the events correspond to the production
of real $Z$-boson in reaction $e^+e^- \rightarrow \gamma Z \rightarrow \gamma
\nu \bar{\nu}$, that is why bounds of paper \cite{10X} are inapplicable 
to reaction $e^+ e^-  \rightarrow \gamma Z^* \rightarrow \gamma N \bar{N}$
for $m_N > m_Z/2$.

For particles with masses of the order of $m_Z/2$ oblique corrections
drastically differ from what we have for masses $\ga m_Z$. In particular,
renormalization of $Z$-boson wave function produces large negative
contribution to $V_A$. From the analysis of the initial set of precision data
in papers \cite{2X, 3X} (published in years 1994 - 1995) it was found that
the existence of additional light fermions with masses $\approx 50$ GeV is
allowed.  Now analyzing all precision data
and using bounds from direct searches we conclude, that the only  presently
allowed light fermion is neutral lepton $N$.  As an example we take $m_U =
220$ GeV, $m_D = 200$ GeV, $m_E = 100$ GeV and draw exclusion plot in
coordinates ($m_N, N_g$), see Fig.3 (we use small $m_U-m_D$ for fitting
purposes).  From this  plot it is clear that for the case of fourth
generation with $m_N \approx 50$ GeV description of the data is not worse
than for the Standard Model and that even two new generations with $m_{N_1}
\approx m_{N_2} \approx 50$ GeV are allowed within $1.5 \sigma$.

\section{The case of SUSY}

In this
section we investigate bounds on extra generations which occur in SUSY
extensions. When SUSY particles are heavy they decouple (i.e.  their
contributions to electroweak observables become power suppressed) and the
same standard model exclusion plots shown in Fig.2 and Fig.3 are valid. The
present  lower bounds on the sparticle masses from direct searches leave
mainly this
decoupled domain.  One possible exception is a contribution of the third
generation squark doublet, enhanced by large stop-sbottom splitting. In this
way we get noticeable positive contributions to functions $V_i$  \cite{36,5}.
They may help to compensate negative contributions of degenerate extra
generations. We analyze the simplest case of the absence of $\tilde {t}_L
-\tilde {t}_R $ mixing in Fig.4.  In this figure the case of degenerate extra
generations with common mass $130$ GeV is considered (contributions of
superpartners of new generations to $V_i$ are negligible since new up- and
down- particles are degenerate).  Exclusion plot is presented in coordinates
$(N_{g}, m_{sbottom})$.  We see that with inclusion of SUSY new
heavy generations are also disfavoured.

One can notice that in the differences $\delta V^{q,l}_m-\delta V^{q,l}_R$ and
 $\delta V^{q,l}_A-\delta V^{q,l}_R$   the dependence on the up-down mass
splitting cancels to
a large extent for fermions and for sfermions as well, that is why {\em if}
degenerate doublets are not allowed by experimental data {\em then}
splitting does
not help. In case of fermions contribution of degenerate family is given
by eq.(16) and it is not allowed because of large negative value of
$\Delta^{\prime}V_R$. In case of sfermions the degenerate family decouples
and one can not compensate contribution of fermions.

 Situation
qualitatively changes in case of light chargino and
neutralino. The latter are still not excluded - dedicated search at LEP II
by DELPHI
still allows the existence of such  particles with masses as
low as $45$ GeV if their mass difference is $\approx 1$ GeV (for larger
$\Delta m$ values charged  decay products of  chargino would be observable; 
for
smaller $\Delta m$ chargino would be seen by its ionization) \cite{14X}.
Analytical formulas for corrections to the functions $V_i$ from
quasi degenerate chargino and neutralino were derived and analyzed in
\cite{14XX}. Corrections are big and this allows one to get lower bounds on
masses of chargino and neutralino:  $m_{\chi}>51$ GeV for the case of higgsino
domination and $m_{\chi}>56$ GeV for the case of wino domination at $95\%$
CL.
\footnote{
These bounds on $m_{\chi}$ follow from the global fit to all electroweak
precision data, while those given in \cite{14XX} are slightly different
since only gluon-free observables were used.} 
 Fig.5 demonstrates how presence of chargino-neutralino pair (dominated
by higgsino) with mass $57$ GeV relaxes the bounds shown on Fig.2. 
 We see that one
extra generation of heavy fermions is allowed within $1.5 \sigma$ domain
in case of the light chargino.

\section{Conclusions}

   Inclusion of new generations in Standard Model is not excluded by
precision data if new neutral leptons are rather light having mass of the
order of $50$ GeV (see Fig.3). Mixing of new leptons with leptons from
three known generations should be small, $\theta \leq 10^{-6}$,
to avoid bounds from direct search at LEP II. We can not exclude stability
of one of these new neutrinos; in this case it becomes interesting for
cosmology. If the early universe was charge symmetric annihilation of $ N
\bar{N}$ in primordial plasma leads to the abundance of these particles at
present time $\Omega < 10^{-3}$ (formula
for heavy neutrino abundance in case $m_N << m_Z$
 was obtained in \cite{17X}). If the early
universe was charge asymmetric abundance of relic $N$'s is larger. However
their contribution to mass density of the halo of our galaxy can not be
larger than 0.1 - 0.01 - otherwise they would be already detected in
laboratory searches for dark matter $\cite{17XX}$. Even this small admixture
of $50$ GeV neutrino in the halo of our galaxy can help to explain gamma
background through  $N \bar{N}$ annihilation into $e^+ e^-$ with subsequent
scattering of electrons and positrons on optical photons \cite{17XXX}.

    Concerning SUSY extensions:
if masses of sparticles are of the order of
several hundred GeV or larger their contribution to electroweak
radiative corrections is negligible,
 hence the above statements remain valid. However in
the case of quasi degenerate chargino and neutralino with masses about $60$ GeV
extra generations of heavy fermions appear to be less forbidden than without
SUSY.

In order to experimentally investigate the case of $m_N < 50 GeV$ a special 
post-LEP II run of LEP I $^\prime$ measuring the Z-line shape slightly  above 
the Z-peak is needed. In this way the bound \cite {1X} will be improved.
For $m_N > 50 GeV$ search for the reaction  
$e^+ e^-  \rightarrow \gamma Z^* \rightarrow \gamma N \bar{N}$
with larger statistics than that of
\cite{10X}  and improved systematics is needed. Finally, further
experimental  search for light  chargino and neutralino \cite{14X} is of
interest. These searches could close the existing windows of ``light''
extra particles, or open  a  door into a  realm of New Physics.


\section{Acknowledgements}

We are grateful to V.F.Obraztsov and F.L.Villante for useful discussions.
The support of RFBR by grants 98-02-17453 and 98-02-17372 is acknowledged.

\newpage
\begin{figure}
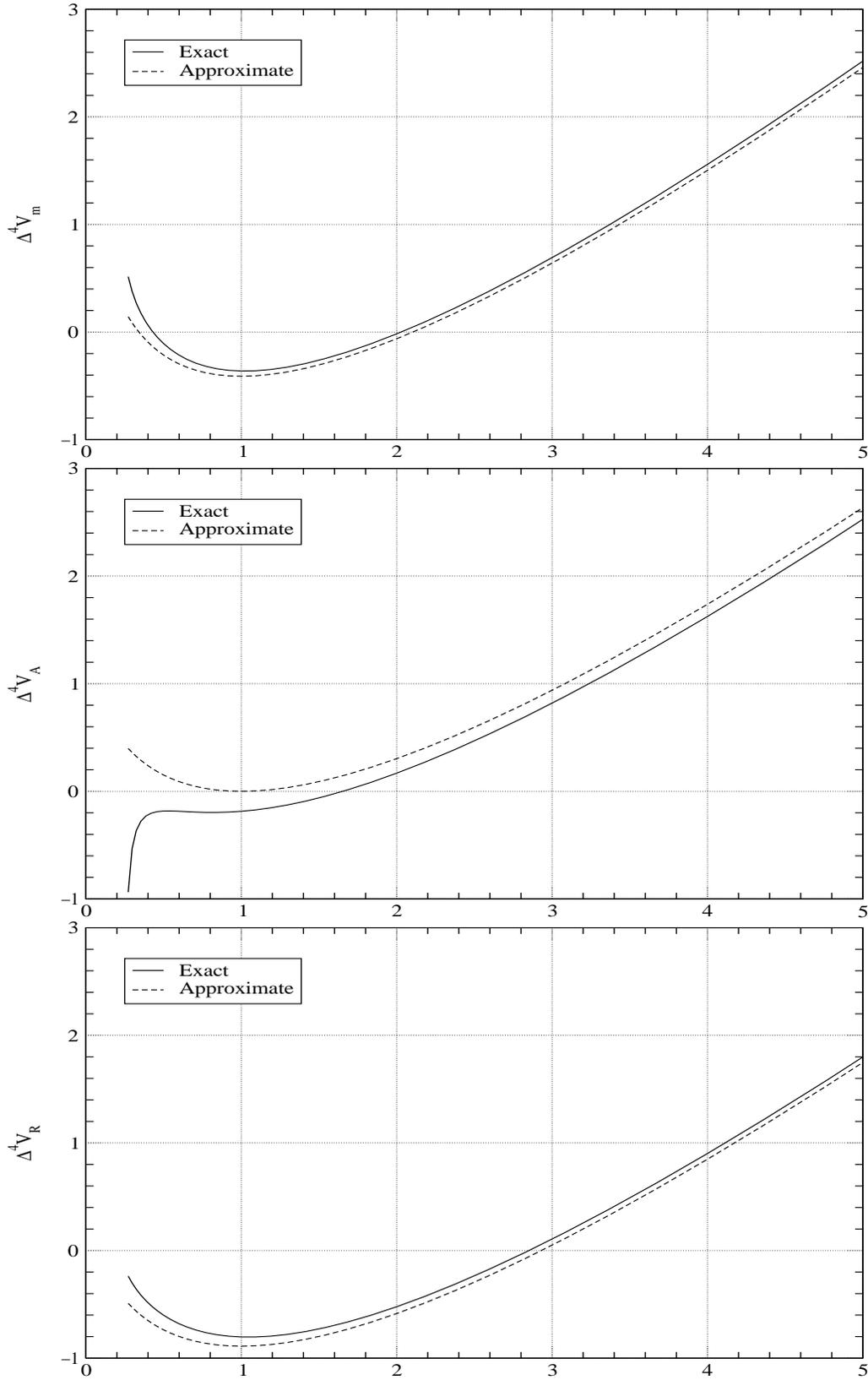
 \centering
\includegraphics[width=0.8\textwidth,height=0.3\textheight]{graph_VM.eps}
\includegraphics[width=0.8\textwidth,height=0.3\textheight]{graph_VA.eps}
\includegraphics[width=0.8\textwidth,height=0.3\textheight]{graph_VR.eps}
\caption{Contributions of a fourth generation of fermions to $\Delta V_i$
as a function of $u \equiv (m_U/m_Z)^2$. We assume $m_N = m_U$ and $m_E =
m_D = m_Z$. Solid lines correspond to exact eq (4);
dashed lines correspond to approximation given
in eq(15). These plots help to study
accuracy of approximation (\ref {7} - \ref {9}) {\rm outside} its formal
domain of validity, that is why we neglect experimental bounds on
$m_D$ and $m_U$.
}
\label{fig1}
\end{figure}

\begin{figure} \centering
\includegraphics[width=0.9\textwidth,height=0.7\textheight]
{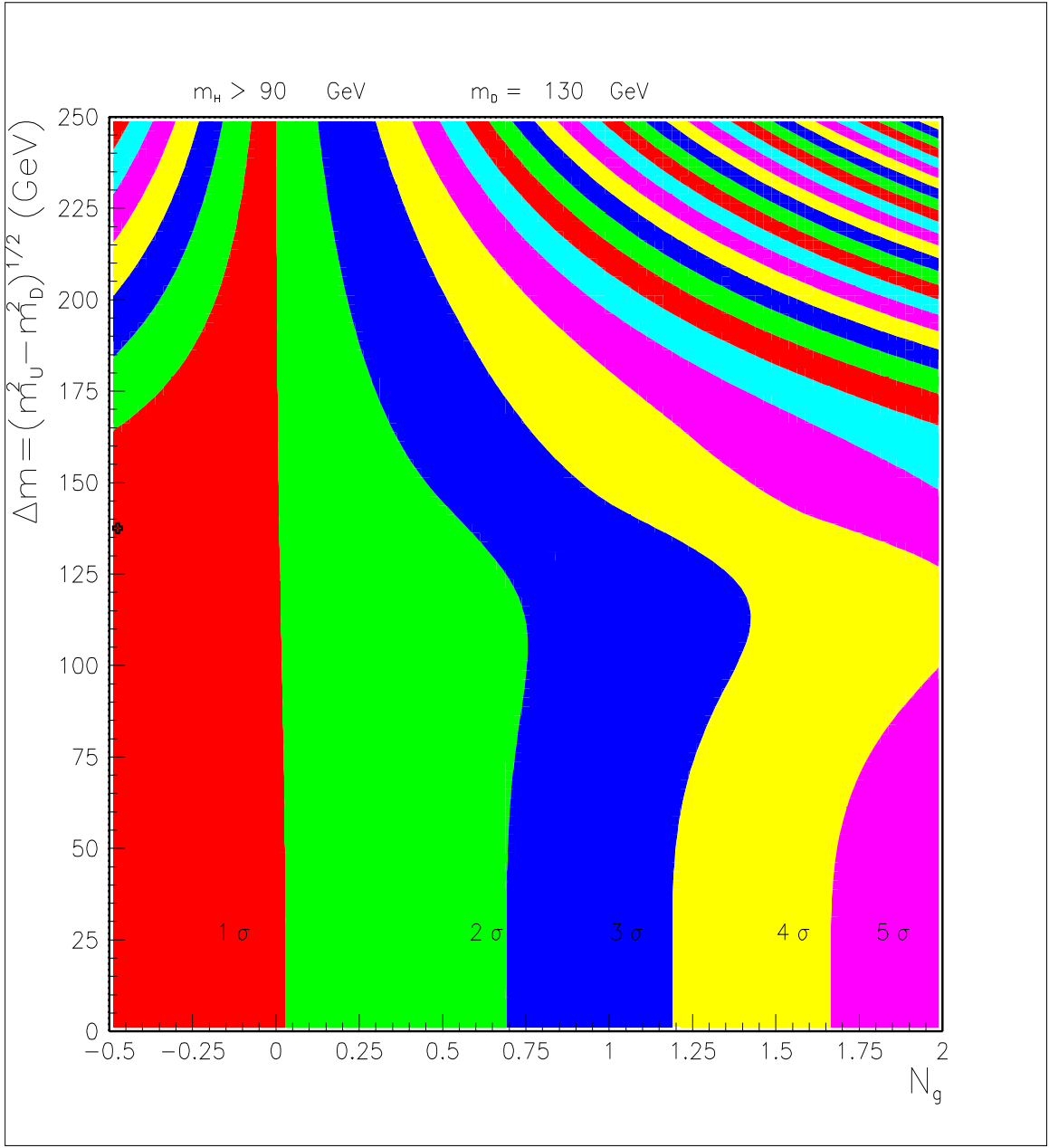}
\caption{
 Constraints on the number of extra generations $N_g$ and the mass
   difference in the extra generations $\Delta m$.
   The lowest allowed value  $m_D=130$ GeV from Tevatron search \cite{13}
   was used and $m_E=m_D$, $m_N=m_U$ was assumed.
   All electroweak precision data and $m_H > 90$ GeV
   at 95 \% C.L. \cite{lep2}  were used in the fit. The cross corresponds to
   $\chi^2$ minimum; regions show  $< 1\sigma$,
$< 2\sigma$, etc. allowed domains.}
\label{fig2}
\end{figure}

\begin{figure} \centering
\includegraphics[width=0.9\textwidth,height=0.7\textheight]{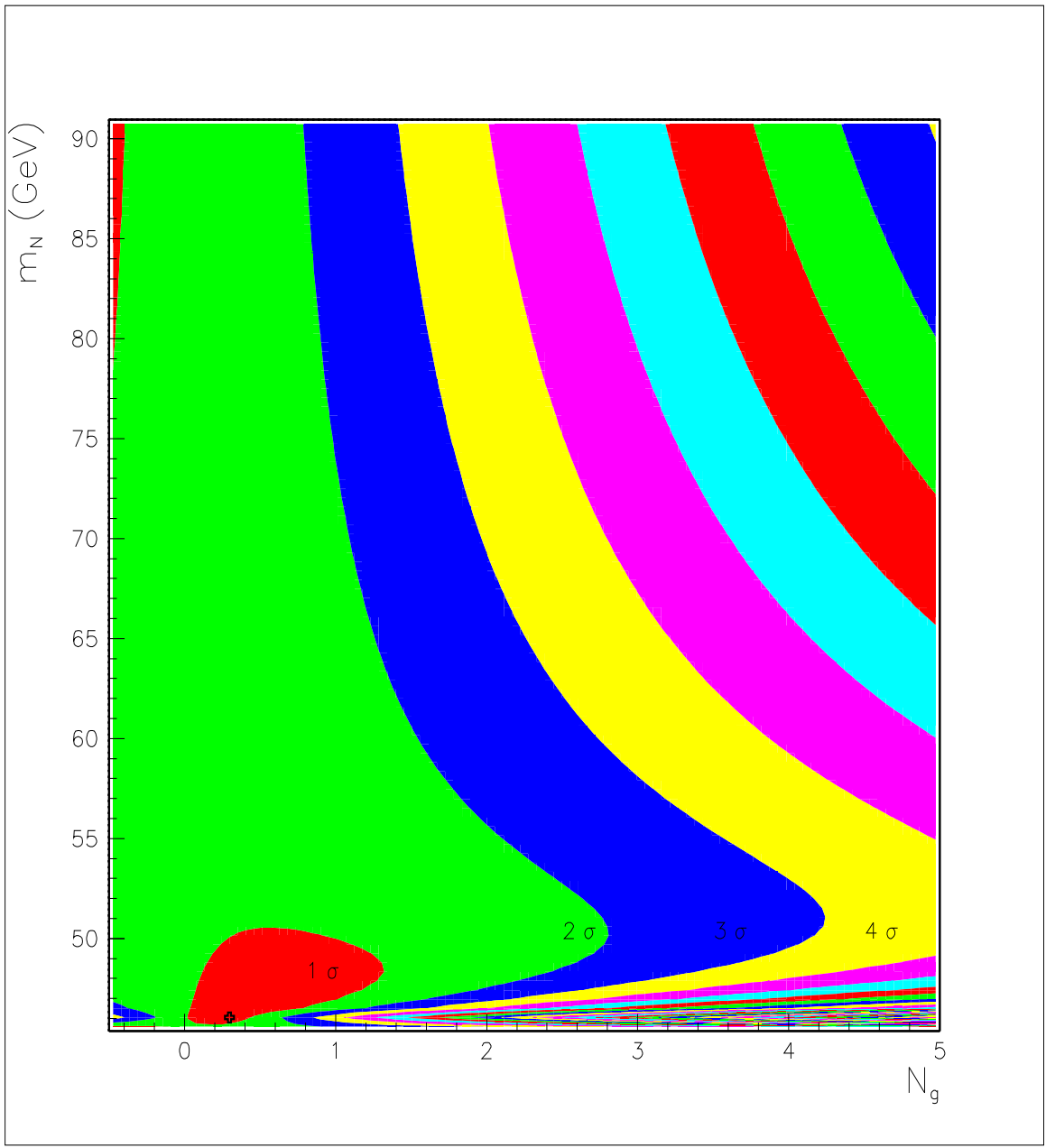}
\caption{
Constraints on the number of extra generations $N_g$ and the mass
   of the neutral heavy lepton $m_N$.
   The values  $m_U=220$ GeV, $m_D=200$ GeV, $M_E=100$ GeV
   were used. All electroweak precision data and $m_H > 90$ GeV
   at 95 \% C.L. from LEP II   \cite{lep2} were used in the fit.
    The cross corresponds to $\chi^2$ minimum; regions show  $< 1\sigma$,
$< 2\sigma$, etc. allowed domains.
}
\label{fig3}
\end{figure}

\begin{figure} \centering
\includegraphics[width=0.9\textwidth,height=0.7\textheight]
{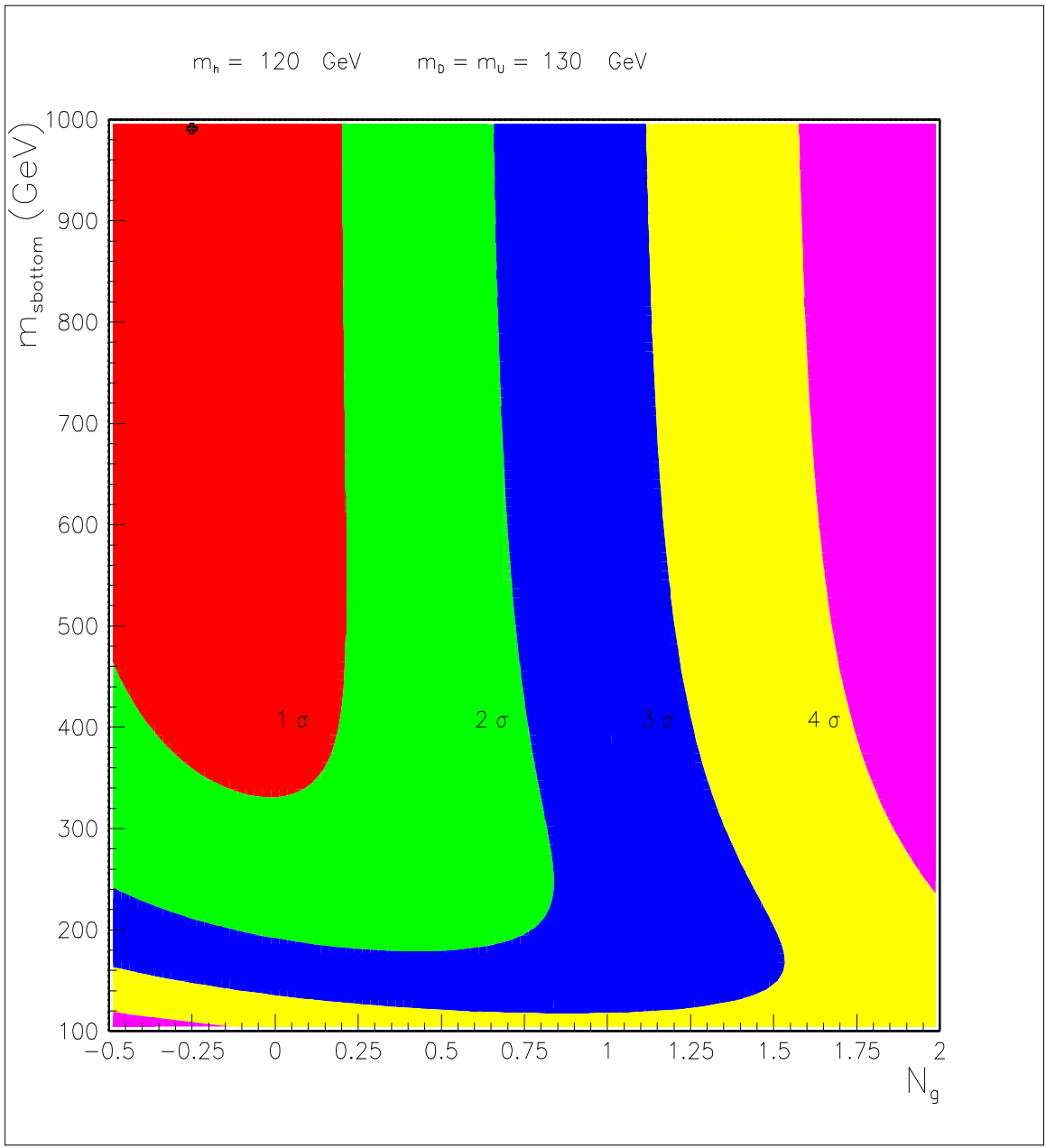}
\caption{
The 2-dimensional exclusion plot for the $N_g$
degenerate extra generations and the mass of sbottom $m_{\tilde{b}}$
in SUSY models and for the choice
  $m_D = m_U = m_E = m_N = 130$ GeV,
using $m_h =120 $ GeV, $m_{\tilde{g}} = 200$ GeV and assuming the absence
of $\tilde {t}_L - \tilde {t}_R$ mixing.
Little cross corresponds to $\chi^2$ minimum; regions show  $< 1\sigma$,
$< 2\sigma$, etc. allowed domains.
}
\label{fig4}
\end{figure}

\begin{figure} \centering
\includegraphics[width=0.9\textwidth,height=0.7\textheight]
{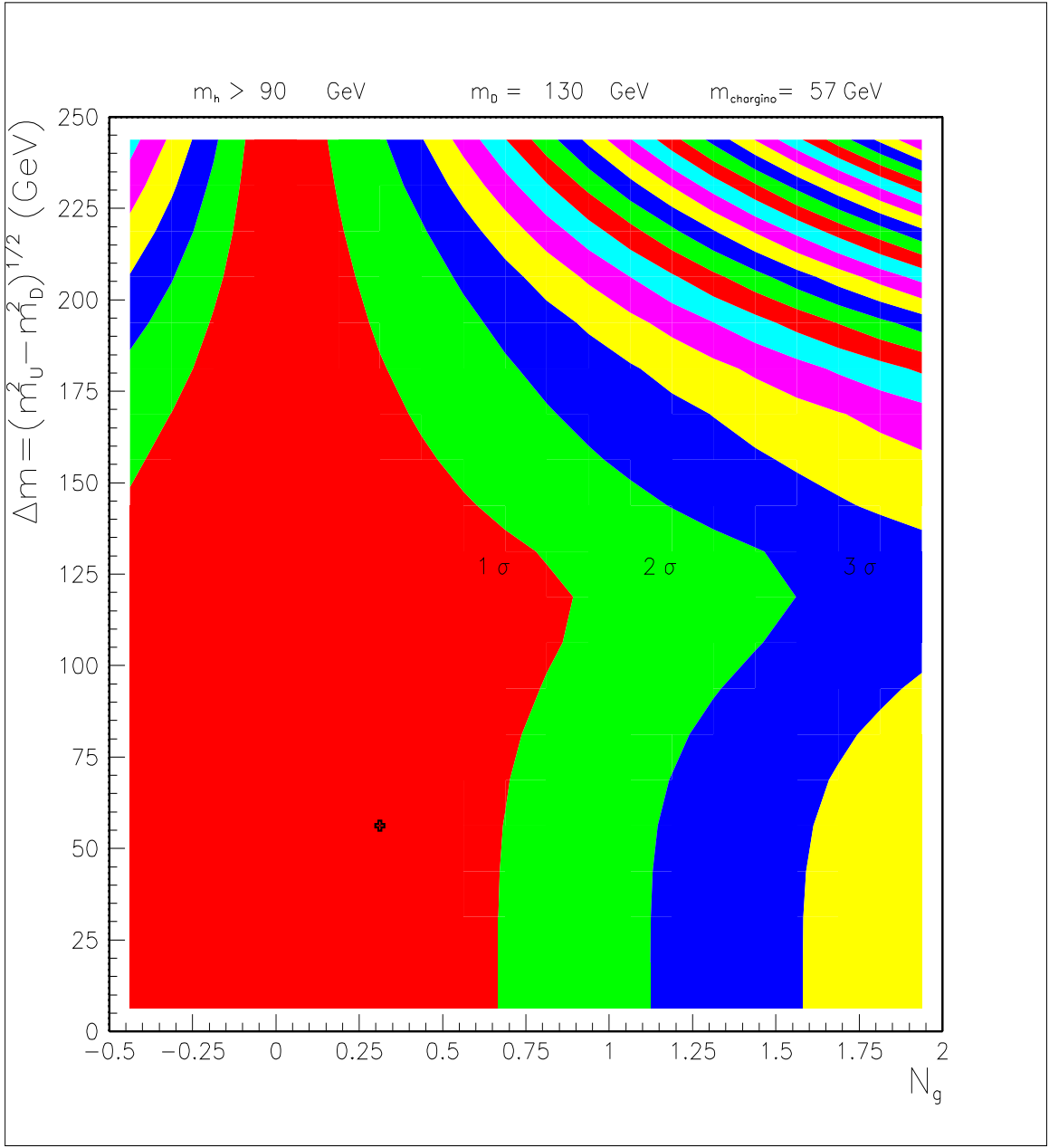}
\caption{
 Constraints on the number of extra generations $N_g$ and the mass
   difference in the extra generations $\Delta m$
in case of $57$ GeV
higgsino-dominated  quasi degenerate chargino and neutralino.
   The lowest allowed value  $m_D=130$ GeV from Tevatron search \cite{13}
   was used and $m_E=m_D$, $m_N=m_U$ was assumed.
   All electroweak precision data and $m_h > 90$ GeV
   at 95 \% C.L. \cite{lep2}  were used in the fit.
    The cross corresponds to $\chi^2$ minimum; regions show  $< 1\sigma$,
$< 2\sigma$, etc. allowed domains.
}
\label{fig5}
\end{figure}
\end{document}